\documentstyle[prc,preprint,aps,epsf]{revtex}

\def\beq{\begin{equation}}
\def\eeq{\end{equation} }
\def\bea{\begin{eqnarray}}
\def\eea{\end{eqnarray}}
\def\eqref#1{Eq.~(\ref{eq:#1})}
\def\eqlab#1{\label{eq:#1}}
\def\figref#1{Fig.~(\ref{fig:#1})}
\def\figlab#1{\label{fig:#1}}

\newcommand{\vslash}[1]{#1 \hspace{-0.5 em} /}
\begin{document}
%\preprint{pinn2}
%\draft
\tighten
\title{Consistent off-shell $\pi N N$ vertex and
nucleon self-energy}
\author{S. Kondratyuk, O. Scholten}
\address{Kernfysisch Versneller Instituut, 9747 AA Groningen, The Netherlands.}
\date{\today}
\maketitle
\begin{abstract}
We present a consistent calculation of half-off-shell form
factors in the pion-nucleon vertex and the nucleon self-energy.
Numerical results are presented.
Near the on-shell point the pion-nucleon vertex is
dominated by the pseudovector coupling, while at
large nucleon invariant masses we find a sizable
pseudoscalar admixture.
\end{abstract}

\vspace{2cm}
\noindent
{\bf Key Words}: Few-body systems, Off-shell form factors, Nucleon
self-energy, Pion-nucleon vertex.

\noindent {\bf 1996 PACS}:
%\pacs
13.75.Gx, 14.20 Dh, 21.45.+v

\bigskip
\noindent
%Corresponding author: \\
%S. Kondratyuk, Kernfysisch Versneller Instituut \\
%Zernikelaan 25, 9747 AA Groningen, The Netherlands  \\
%e-mail:  KONDRAT@KVI.NL, \\
%phone:  +31-(0)50-3636192, fax:    +31-(0)50-3634003 \\

- - - - - - - - - - - - - - - - -  \today\ - - - - - - - - - - - - - - - - -

\pacs{13.40.Gp, 21.45.+v, 24.10.Jv, 25.20.Lj}

%\newpage
\section{Introduction}

The structure of hadronic vertices, usually parametrized in terms of form
factors, is important in much of nuclear physics. The form
factors may depend on the different invariants that can be constructed. In
nucleon-meson or nucleon-photon vertices one often considers only the dependence
on the momentum-squared of the meson or photon. In the present paper we will
consider so-called off-shell form factors where the dependence is studied on the
momentum-squared of one of the nucleons involved.

Off-shell form factors are an ingredient in the description of physical
processes. For example, nucleon-photon off-shell form factors have been shown to be
important in models for proton-proton bremsstrahlung
\cite{nyman,kondrsch,li} and virtual Compton scattering\cite{kor}. $\pi N N$ and other
nucleon-meson form factors with
an explicit dependence on the momentum of one or both nucleons have been used in models for
$N N$ \cite{grossord} and $\pi N$ \cite{gross,pearce,julich,sato} scattering, pion photoproduction
\cite{sato}
and vector meson production in nucleon-nucleon collisions \cite{naka}. In these models,
the form factors have been phenomenologically parametrized,
with the parameters adjusted to fit experimental data.

The off-shell structure of the nucleon-photon vertex\cite{bincer,photoffdis,photoffloop},
and the nucleon-pion vertex\cite{bincer,pionoffdis,pionoffnutt} has been studied before.
In particular, dispersion relation techniques are used in
Refs. \cite{bincer,photoffdis,pionoffdis}, whereas the models of Refs. \cite{photoffloop} are
based on a perturbative dressing of the vertex with one-meson loops.
In this work we investigate the
pion-nucleon coupling in a field-theoretical model which is inherently
non-perturbative and is based on the Schwinger-Dyson equation, considering
loops to all orders. The nucleon self-energy and the pion-nucleon vertex
function are both calculated in a consistent framework.

In general, off-shell form factors and the functions parametrizing the
self-energy are complex functions, where the imaginary parts are
related to open multi-particle chanels of which the pion-nucleon
channel will be the most important.
Our approach is based on the analytic structure\cite{bogol,bincer}
of the nucleon self-energy  and the off-shell $\pi N N$
vertex, which is exploited by the use of dispersion relations. The
imaginary parts of the form factors and the self-energy are calculated from Cutkosky
rules\cite{cut}. To make this procedure tractable, we consistently neglect
contributions to the imaginary parts from the multi-pion thresholds.

For the course of this paper, we are interested in vertices with one
off-shell nucleon, which contain two independent form factors. The
convergence of the loop corrections is insured through the introduction of a
``cut-off" function (the initial form factor). We
obtained the interesting result that the widths of the converged form factors
have an upper bound.

Solutions of the Schwinger-Dyson equation have been presented in the past
(see, e.g.,\cite{brown,krein} and a recent paper\cite{wilets}). There, a
usually adopted approximation consists in assuming the same spin structure for the
dressed and the bare vertex. The dressing of
the vertex is thus parametrized in terms of a single form factor. In the
present work we have released this condition and found a strong dependence
of the spin structure of the vertex on the off-shellness involved.

Form factors are usually interpreted as representing the features that are
not included explicitely in a particular model for a physical process, and
as such, should be built consistently with the kind of models in which they
are intended to be used. The form factors considered in the present paper
are primarily designed for usage in a K-matrix model for pion-nucleon
scattering, pion photoproduction and Compton scattering off the nucleon
\cite{gouds,korsch}. Since in such a model the one-pion production channel is
included explicitely, only real form factor should be used there (at least
below the two-pion threshold). This aspect is elaborated on in Section
\ref{sec:Kmat}.

In any model, treatment of off-shell three-point vertices should be
linked with treatment of higher-point vertices, because
a redefinition of the nucleon field can change off-shell dependence of the
former in favour of presence of the latter. The
observables are oblivious to the representation of fields
(this result is known as the equivalence theorem)\cite{kamef},
examples of which can be found, e.g., in Refs.\cite{scherfear,davpoul,grossoff}.
In the present model, higher-point vertices are excluded at all stages of the
calculations, and the discussion is carried out solely in terms of
off-shell form factors in the $\pi N N$ vertex.

The paper is organized as follows. In Section II the general structure
of the off-shell $\pi N N$ vertex is discussed. Our model is described in detail in
Section III. At present we limit ourselves to the inclusion of
one-pion-nucleon loops only, for which numerical results are presented in
Section IV.%\ref{sec:III}.

\section{Structure of the $\pi N N$ vertex\label{sec:Struc}}

The $\pi N N$ vertex operator is the sum of all connected Feynman diagrams
with one incoming nucleon (carrying the momentum $p$), one outgoing nucleon
($p^\prime$) and one pion ($q=p-p^\prime$), with the propagators for
the external legs stripped away. The most general form compatible with
Lorentz covariance and isospin invariance  reads\cite{kazes}
\begin{eqnarray}
\Gamma_{\alpha}(p^{\prime},p,q) & = &
\tau_{\alpha}\Big(\gamma^5 G_1(p^{\prime 2},p^2,q^2)+
\gamma^5 \frac{\vslash{p}-m}{m} G_2(p^{\prime 2},p^2,q^2) +
\frac{\vslash{p}^\prime-m}{m} \gamma^5 G_3(p^{\prime 2},p^2,q^2) \nonumber\\
 & & +\frac{\vslash{p}^\prime-m}{m} \gamma^5 \frac{\vslash{p}-m}{m}
G_4(p^{\prime 2},p^2,q^2)\Big),
\eqlab{GenV}
\end{eqnarray}
where $m$ denotes the nucleon mass and $\tau_{\alpha}, \alpha=1,2,3,$ are the
isospin Pauli matrices. The form
factors $G_i$ depend on the three Lorentz scalars,
$p^{\prime 2}, p^2$ and $q^2$. Usually the situation is considered in which both
nucleons are on the mass shell, i.e. $p^{\prime 2}=p^2=m^2$, and only $G_1(m^2,m^2,q^2)$
enters in \eqref{GenV}. In this paper we consider a different situation in
which the pion and only one of the nucleons is
on the respective mass shell, $p^{\prime 2}=m^2$ and $q^2=\mu^2$, where $\mu$
denotes the pion mass. Such a vertex is
conventionally called the half-off-shell $\pi N N$ vertex, and
it contains so-called half-off-shell form factors.

If the operator of \eqref{GenV} works on the positive energy spinor
$\overline{u}(p^{\prime})$ to the left, the last two terms in \eqref{GenV}
vanish due to the Dirac equation,
$\overline{u}(p^{\prime})\,\vslash{p}^{\prime}=\overline{u}(p^{\prime})\,m$,
and the vertex contains only the form factors
$G_1(m^2,p^2,\mu^2)$ and $G_2(m^2,p^2,\mu^2)$.
Similarly, if the initial nucleon is on-shell, only the form
factors $G_1(p^{\prime 2},m^2,\mu^2)$ and $G_3(p^{\prime 2},m^2,\mu^2)$
are left. Charge-conjugation, space-inversion and time-reversal
symmetries allow to relate these form factors,
\begin{equation}
G_1(p^2,m^2,\mu^2)  =  G_1(m^2,p^2,\mu^2),\;\;\;\;\;\;\;
G_3(p^2,m^2,\mu^2)  =  G_2(m^2,p^2,\mu^2).
\eqlab{2}
\end{equation}
Hence, one can consider only the vertex with the outgoing on-shell nucleon.
Omitting the trivial arguments in $G_i$,
\begin{equation}
\Gamma_{\alpha}(m,p,\mu)=\tau_{\alpha}\, \Gamma (m,p,\mu) =
\tau_{\alpha}\,\gamma^5 \Big(G_1(p^2)+ \frac{\vslash{p}-m}{m} G_2(p^2)\Big),
\eqlab{3}
\end{equation}
where the notation $\Gamma_{\alpha}(m,p,\mu)$ implies that $q^2=\mu^2$ and
$\overline{u}(p^{\prime})\,\vslash{p}^{\prime}=\overline{u}(p^{\prime})\,m$
in all expressions for this vertex.
Along with \eqref{3}, we will use another form for the half-off-shell vertex,
\begin{equation}
\Gamma_{\alpha}(m,p,\mu)=
\tau_{\alpha}\,\gamma^5 \Big( G_{PS}(p^2) + \frac{\vslash{p}+m}{2m}
G_{PV}(p^2)\Big),
\eqlab{4}
\end{equation}
where
\begin{equation}
G_{PS}(p^2)=G_1(p^2)-2 G_2(p^2) , \;\;\;\;\;\;\;\; G_{PV}(p^2)=2 G_2(p^2)\,,
\eqlab{5}
\end{equation}
denote the form factors corresponding to the usual pseudoscalar and
pseudovector couplings.

\section{Description of the model \label{sec:II}}

The model for the form factors is based on a non-perturbative dressing of
the vertex with pion loops as represented graphically in Fig. 1. The nucleon
self-energy $\Sigma(p)$ is calculated self-consistently using the
Schwinger-Dyson equation\cite{landau} with the dressed vertex. This can be
expressed in terms of a system of integral equations for the dressed
$\pi N N$ vertex $\Gamma_{\alpha}(p^{\prime},p,q)$ and the dressed nucleon
propagator $S(p)$,
\beq
 \left\{
\begin{array}{ccl}
\Gamma_{\alpha}(m,p,\mu)  & = &\Gamma_{\alpha}^0(m,p,\mu)-
i{\displaystyle \int \frac{d^4 k}{(2\pi)^4}}\,
\Big( \Gamma_{\beta}(m,p^{\prime}+k,k)\, S(p^{\prime}+k)  \\
 & &  \times\Gamma_{\alpha}(p^{\prime}+k,p+k,\mu)\, S(p+k)\,
\Gamma_{\beta}(p+k,p,-k)\,D(k^2) \Big),
\\
S(p)&=&S^0(p)+S(p)\,\Sigma(p) \, S^0(p)  ,\\
\Sigma(p) &=& -i{\displaystyle \int\frac{d^4 k}{(2\pi)^4}} \,\Big(
 \Gamma_{\alpha}(p,p+k,k) \,
S(p+k)\, \Gamma_{\alpha}^0(p+k,p,-k)\, D(k^2)\Big)  \\
 & & -(Z_2-1)(\vslash{p}-m)-Z_2\,\delta m,
%\hfill (11)
\end{array} \right.
\eqlab{sys} \eeq
where $D$ is the pion propagator, $S^0$ the free propagator of the nucleon
and $\Gamma_{\alpha}^0(m,p,\mu)$ the bare $\pi N N$ vertex. The last two
terms in the equation for the self-energy are part of the renormalization
procedure and will be discussed later. The dressing is non-perturbative
since the dressed vertex and propagator appear also on the right-hand side
of the equations. As is well known, such a
procedure suffers from divergences. In addition, we are interested to build
a model for half-off-shell vertices while the right-hand sides of
Eqs.~(\ref{eq:sys}) include vertices with all external legs off shell. To
circumvent these two problems we have applied a solution procedure based on
the use of dispersion relations and a regularization method as outlined in
the following sections.

 \subsection{Solution procedure }

Hereafter we shall denote half-off-shell vertices as
$\Gamma_{\alpha}(p)$, dropping the trivial parameters for brevity. As
noted in the above, Eqs.~(\ref{eq:sys}) require the knowledge of the
full off-shell vertex. Using the analyticity of the form factors and the
self-energy this problem can be bypassed. The imaginary parts of the form
factors can be obtained by applying Cutkosky rules to the integrals in
Eqs.~(\ref{eq:sys}), and we reconstruct the real parts through the application of
dispersion relations implemented in an iterative procedure.

As stated in the introduction,
we have excluded four- or
higher-point vertices from our model. Furthermore,
in applying Cutkosky rules we shall only include the channel with the lowest
threshold, i.e.\ the one-pion nucleon channel. This results in the
contributions depicted in Fig.~2.

The solution procedure can now be explained best by going in some
detail through one complete step of the iterative procedure. From the
n$^{\rm th}$ iteration we have obtained the form factors $G^n_{1,2}(p^2)$
which define the vertex through \eqref{3} and the self-energy functions
$A^n(p^2)$ and $B^n(p^2)$ which define the dressed propagator,
\beq
\big( S^{n}(p) \big)^{-1} = Z_2(\vslash{p}-m_B)-
\Big[ A^n(p^2) \vslash{p} + B^n(p^2) m \Big] ,
\eqlab{Sdrs}
\eeq
where the term in square brackets is the loop contribution
$\Sigma^{n}_L(p)$ to the
self-energy. The parameters $Z_2$ and $m_B=m-\delta m$ are
renormalization constants as defined in Sec.~(\ref{Ren}).

The imaginary parts of the
form factors and the self-energy functions arise from the pinching-pole
term in the loop integrals which can be evaluated using Cutkosky
rules\cite{cut}. This contribution is labelled by the subscript $I$.
For the self-energy one has:
\beq
\Sigma_I^{n+1}(p) = -\overline{\Gamma}^{n}_{\alpha,R}(p)\, I_{pole}(p)\,
\Gamma^{n}_{\alpha,R}(p)\,, \eqlab{SnI}
\eeq
where the subscript $R$ denotes the vertex
calculated using only the real parts of the form factors $G_{1,2}^n$.
The Dirac conjugated vertex is denoted as
$\overline{\Gamma}$.
The remaining integral can be written as
\beq
I_{pole}(p) = \frac{1}{8\pi^2} \int\!d^4 k\, (\vslash{p}+\vslash{k}+m)\,
\delta((p+k)^2-m^2)\,\theta(p_0+k_0)\,\delta(k^2-\mu^2)\,\theta(-k_0)\, .
\eqlab{InPole}
\eeq
The explicit form for $I_{pole}(p)$ is given in Appendix A.
The imaginary parts of
$A^{n+1}(p^2),B^{n+1}(p^2)$ can now readily be written
using \eqref{Sdrs}.

The real parts of the self-energy functions are calculated via the
dispersion relations\cite{bogol},
\beq
Re\,A^{n+1}(p^2) = \frac{\mathcal{P}}{\pi} \int_{w_{th}^2}^{\infty} \!\!
   dp^{\prime 2}\, \frac{Im\,A^{n+1}(p^{\prime 2})}{p^{\prime 2}-p^2}\,,
\eqlab{dispA}
\eeq
and similar for $Re\,B(p^2)$.
Here $w_{th} \equiv (m+\mu)$ is the one-pion threshold,
and $\mathcal{P}$ denotes the principal value integral.

The pole term in the loop integral for the vertex reads:
\beq
\Gamma^{n+1}_{\alpha,I}(p) =
J_{pole}(p)\, \Gamma^{n}_{\alpha,R}(p)\,, \eqlab{GnI}
\eeq
with
\bea
J_{pole}(p) &=& -\frac{(-2)}{8\pi^2}\,\int\!d^4 k\, \Gamma^{n}_{R}(p'+k) \,
S^{n+1}_R(p'+k) \, \overline{\Gamma}^{n}_{R}(p'+k) \,
(\vslash{p}+\vslash{k}+m)\,
\nonumber \\ & & \times \delta((p+k)^2-m^2)\, \theta(p_0+k_0)\,
\delta(k^2-\mu^2)\, \theta(-k_0)\,, \eqlab{JnPole}
\eea
where the integral \eqref{JnPole} is independent of $p'$, the momentum of
the outgoing on-shell nucleon, as shown in Appendix B. The factor ($-2$) in
\eqref{JnPole} comes from commuting the isospin matrices. The propagator
$S^{n+1}_R$ in \eqref{JnPole} contains only the real parts of
$A^{n+1}$ and $B^{n+1}$ which have been calculated by virtue of
Eqs.~(\ref{eq:SnI},\ref{eq:InPole},\ref{eq:dispA}).
Casting $\Gamma^{n+1}_{\alpha,I}(p)$ in the form of \eqref{3},
the imaginary parts of the
form factors $G_{1,2}^{n+1}(p^2)$
are found.

To construct the real parts of the form factors
we take advantage of their analytical properties\cite{bincer},
\beq
Re\,G^{n+1}_i(p^2) = G^0_i(p^2)+ \frac{\mathcal{P}}{\pi}
      \int_{w_{th}^2}^{\infty} \!\! dp^{\prime 2}\,
    \frac{Im\,G^{n+1}_i(p^{\prime 2})}{p^{\prime 2}-p^2}\,. \eqlab{dispG}
\eeq
The first term on the right-hand side of \eqref{dispG} derives from the
equivalent term in Eqs.~(\ref{eq:sys}).
We use unsubtracted dispersion relations since convergence of the
integrals can be guaranteed by the cut-off function introduced in
\eqref{Flam}.

There are a few points that need special stressing here.
\begin{itemize}

\item In calculating the imaginary parts for
the (n+1)$^{\rm st}$ iteration, we retain only the real parts of the form
factors and self-energy functions from the n$^{th}$ iteration, to be
consistent with the use in a K-matrix formalism as explained in
Sec.~(\ref{sec:Kmat}).

\item Except the depicted cuts, any other kinematically allowed cuts of the
loop diagrams (cutting through the blobs in \figref{f2}) would correspond to either picking
up contributions of higher thresholds or considering four-point vertices,
both of which would be inconsistent with the adopted solution scheme.

\item In cutting the self-energy loop diagram, dressed vertices at both
sides of the cut propagator are taken into account. Seemingly, this is in
conflict with the Schwinger-Dyson equation (see Eqs.~(\ref{eq:sys}) and Fig. 1),
where the second blob would lead to double counting. However, the presence
of the two blobs in the {\it cut} diagram is necessary to sum up {\it all}
contributions from one-pion-nucleon cuts.

\item The present method of solution allows one to avoid dealing with the
full off-shell vertices present Eqs.~(\ref{eq:sys}). Indeed, as can be seen from
Fig. 2, we need only half-off-shell vertices throughout the iteration
process.

\item To calculate the pole contributions of the loop integrals we have
applied Cutkosky rules i.e.\ put the nucleon and pion lines in the loop
integrals on their respective mass shells, as shown in Fig. 2. In
the cut propagators, therefore, only physical masses appear. In particular,
this implies that the dressing of the pion propagator does not have to be
considered in the present approach.

\end{itemize}

We take
$\Gamma_{\alpha}^0(m,p,\mu)$ as zeroth iteration for the vertex (its precise
form is specified in \eqref{Flam}). At each
iteration step, we utilize the dispersion relations
Eqs.~(\ref{eq:dispG},\ref{eq:dispA}), where $Im\,G_{1,2}(p^2),\,Im\,A(p^2)$
and $Im\,B(p^2)$ are calculated using Cutkosky rules. In this
connection, the following remarks are in order.

{\it Analyticity.} \\ In principle, the use of the dispersion relations
should guarantee that the form factors and the self-energy functions $A$ and
$B$ be analytic in the complex plane cut from $w_{th}^2$
to $\infty$ along the real axis. However, the actual imagianary parts
calculated in the model contain also ``unphysical" singularities of the
function $G^0(p^2)$, see \eqref{Flam}, regularizing the dispersion integrals.
This, strictly speaking, invalidates the derivation of the
dispersion relations. This problem will be encountered for any non-constant
function $G^0(p^2)$ unless its singularities are located along the cut.

{\it Unitarity} \\ In applying Cutkosky rules, an important qualification is
that we neglect the contributions to the imaginary parts that come from the
intermediate states including one nucleon and more than one pion. In order
that unitarity should hold exactly, the imaginary parts must contain the
contributions from all multi-pion thresholds. In the context of the present
work, a rigorous account of, e.g., the two-pion threshold would require
computing the imaginary parts of two-loop self-energy and vertex diagrams.
Analyses of massive two-loop Feynman diagrams have appeared in the
literature recently, including the dispersion relation approach (see, e.g.,
\cite{bautausk} and references therein). However, such calculations are
rather complicated, and we
found their application in the present model not feasible.

One may argue that the real function $G^0$, which is used presently, could
be expressed in terms of a dispersion integral over a function, say $F^0$,
which would correspond to the discontinuities of the form factors due to all channels
opening at higher thresholds not considered explicitely. If this were the case,
the presently adopted procedure would be equivalent to adding an extra
contribution $F^0$ to the imaginary parts derived from the loop diagrams.
This could possibly account
for the two problems just mentioned, but we did not pursue this direction.

\subsection{Renormalization and regularization\label{Ren}}

The renormalized nucleon self-energy in Eqs.~(\ref{eq:sys}) or \eqref{Sdrs} can
be written as
\beq
\Sigma(p) = \Sigma_L(p) -(Z_2-1)(\vslash{p}-m)-Z_2\,\delta m.
\eqlab{12}
\eeq
The first term in \eqref{12} is the contribution of pion
loops while the last two terms come from the counterterms in
the Lagrangian as part of the renormalization procedure.

The construction of the counterterms is based on the usual
renormalization procedure\cite{weinb} as explained by the following example.
The Lagrangian, written in terms of the ``bare" fields,
masses and coupling constant, is
\beq
{\mathcal L} = \frac{1}{2}
 ({\partial}_{\nu}{\phi}_B\,{\partial}^{\nu}{\phi}_B- {\mu}_B^2\,{\phi}_B^2)
 + {\overline{\psi}}_B(i\,/\!\!\!{\partial} - m_B){\psi}_B
 - \frac{g_B}{2m_B}\, \overline{\psi}_B \gamma^5\,
     (/\!\!\!{\partial}{\phi}_B) {\psi}_B.
\eqlab{bare}
\eeq
Defining the renormalized nucleon field $\psi=Z_2^{-1/2}\,{\psi}_B$, the
renormalized nucleon mass $m=m_B+\delta m$ and the
constant $f/(2m)=g_B Z_2/(2 m_B)$, \eqref{bare} can
be reformulated as
\begin{eqnarray}
\mathcal{L} &=& \frac{1}{2}({\partial}_{\nu}{\phi}\,{\partial}^{\nu}{\phi}-
{\mu}^2\,{\phi}^2) + {\overline{\psi}}(i\,/\!\!\!{\partial}-m){\psi}
\nonumber \\
 & & -\Big(\frac{f}{2m}\,\overline{\psi}\gamma^5\,
(/\!\!\!{\partial}{\phi}){\psi}-
Z_2\,\delta m\,\overline{\psi}\psi-(Z_2-1)\,\overline{\psi}
(i\,/\!\!\!{\partial}-m){\psi}\Big).
\eqlab{bare-ren}
\end{eqnarray}
Because we encounter only cut pion lines during the iteration procedure
(see Fig.~2), we need only the pole contribution of the pion propagator.
Since the pole properties of the renormalized dressed propagator coincide
with those of the free one,
the pion field and mass need not be renormalized in our approach,
$\phi=\phi_B, \mu^2=\mu^2_B$. From \eqref{Sdrs} and \eqref{12} it can be
seen that the renormalization constants $Z_2$ and $\delta m$
can also be interpreted as
real constants which can always be added when the real part of a function is determined
from the imaginary part via a dispersion relation \eqref{dispA}.

The coupling strength $f$ and the renormalization constants
$Z_2, \delta m$ are determined by fixing $\Gamma_{\alpha}(m,m,\mu)$ and
the pole structure of the propagator $S(p)$,
\bea
&&S^{-1}(m)=0,\;\;\;\;\;\; \nonumber\\
&& Res[S(p),\,\vslash{p}=m]=1, \eqlab{8} \\
&&\overline{u}(p^{\prime})\,\Gamma_{\alpha}(m,m,\mu)\,u(p)=
\overline{u}(p^{\prime})\,\tau_{\alpha}\,\gamma^5\,g\,u(p), \nonumber
\eea
where the last equation can be reduced to $G_1(m^2)=g$, the
physical pion-nucleon coupling constant (we take $g=13.02$\cite{rob}).
The left-hand side of this condition is calculated at the kinematically
forbidden point, where all the external legs of the vertex are on-shell.
However, this is of no harm for the renormalization prescription.
We could choose any convenient renormalization point as long as the
form factors calculated at that point are real (see, e.g.\ \cite{weinb},
where the freedom of the choice of a renormalization procedure
is discussed in general).

In the context of the iterative procedure described in the previous section,
the constants $Z_2$ and $\delta m$ are chosen to provide the
correct pole properties of the {\it converged} propagator.
This implies that the pole location and residue of the propagator
are off in the course of the first few iterations. To check that
this feature is immaterial for the final result, we applied also another
solution procedure. Its main difference from the one outlined above
is that the renormalization of the propagator is done at each
iteration step, insuring the correct pole properties
at any iteration. We found that both methods lead to identical
results for the converged vertex and propagator.
The reason for this is that when convergence has been reached, a
non-perturbative solution of Eqs.~(\ref{eq:sys}) (under the provisions which have
been discussed) has been obtained. The intermediate steps in the iteration
procedure at this point are an uninteresting technical detail.

The loop integrals, or
rather the dispersion integrals Eqs.~(\ref{eq:dispG},\ref{eq:dispA}),
will diverge unless a regularization is applied.
As part of the regularization procedure, we introduced a
form factor $G^0(p^2)$ (also called the cut-off function) for the bare
$\pi N N$ vertex,
\beq
\Gamma_{\alpha}^0(m,p,\mu)=\tau_{\alpha}\gamma^5
\frac{\vslash{p}+m}{2 m}\, G^0 (p^2),
\eqlab{Flam}
\eeq
in terms of \eqref{4}.
The function $G^0 (p^2) \equiv G_{PV}^0(p^2)$ is normalized to $f$ at $p^2=m^2$
and must fall off sufficiently fast at infinity to provide convergence of
the integrals. In the numerical example discussed later we used two different
functions $G^0 (p^2)$, see Eqs.~(\ref{eq:23},\ref{eq:24}).

The cut-off
function is a phenomenological input of the model.
A self-consistent
procedure to construct meson-nucleon form factors was presented in Ref.\ \cite{flender},
where both nucleons in the vertex are on-shell
and the meson is off-shell. There, no phenomenological form factor
was needed. We were not successful in implementing a similar approach
for half-off-shell form factors in the pion-nucleon vertex,
nontrivial solutions of the relevant equations for the self-energy and
the $\pi N N$ vertex did not seem to exist.

\subsection{Consistency with a K-matrix approach\label{sec:Kmat}}

One motivation for the present model is the construction of form factors and
self-energies which can be applied in a K-matrix approach to $\pi N$ scattering
\cite{gouds,korsch}. We outline the K-matrix method (details
can be found in\cite{gouds}) and in particular address the double-counting
issue: by considering only the real part of the form factors and the
self-energy functions
on the r.h.s. of Eqs.~(\ref{eq:SnI},\ref{eq:GnI},\ref{eq:JnPole}),
we avoid double
counting when the calcualted vertex and propagator are used in the
K-matrix approach.
It should be emphasized that only the one-pion threshold
discontinuities are taken into account in both the K-matrix approach in
question\cite{gouds,korsch} and the present model.

The Bethe-Salpeter equation for the $\pi N$ scattering amplitude $\mathcal{T}$
can be written in the operator form,
\beq
{\mathcal{T}}\,=\,V\, +\,V\,{\mathcal G}\,{\mathcal T}\,.
\eqlab{k1}
\eeq
Here, $V$ is the sum of all irreducible diagrams describing the
scattering, and $\mathcal {G}$ is the free $\pi N$ propagator.
$\mathcal {G}$ can be decomposed as the sum of the on-shell
contribution $i\delta$ which is
imaginary (according to Cutkosky rules), and the off-shell part
$\mathcal {G}^P$ which is real,
\beq
{\mathcal G}\,=\,{\mathcal G}^P\,+\,i\delta,
\eqlab{k2}
\eeq
where $\delta$ implies that the
corresponding intermediate nucleon and pion are taken on their respective
mass shells. The K-operator is introduced by the equation
\beq
K\,=\,V\,+\,V\,{\mathcal G}^P\,K\,.
\eqlab{k3}
\eeq
Combining the last three equations yields the ${\mathcal T}$-matrix
expressed in terms of the K-matrix,
\beq
{\mathcal{T}}\,=\,K\,+\,K\,i\delta\,{\mathcal{T}}\, .
\eqlab{k4}
\eeq
This  is the central equation used in a K-matrix approach and can
schematically be written as ${\mathcal T}= K / (1-i K)$.
If $K$ is hermitian, the scattering operator $S=1+2i{\mathcal{T}}$ will
be unitary.

We consider a simplified version of the K-matrix approach containing only
nucleons and pions, with the kernel $V$ chosen as the sum of the s- and
u-channel tree diagrams. One way to construct the K-operator is to set $K=V$
\cite{korsch}, thereby assuming ${\mathcal{G}}^P=0$, see \eqref{k3}. Then,
by \eqref{k4}, the $\mathcal{T}$-matrix will contain the loop diagrams in
which only the cut nucleon and pion propagators will enter. Only by
using dressed vertices and propagators one may
take the K-matrix equal to the sum of skeleton diagrams solely.
As implied by \eqref{k3}, the
form factors in these dressed vertices take into account real contributions
due to the principal
value ${\mathcal G}^P$. These are the real parts of the form factors
discussed in the previous sections. If we kept both the real and
imaginary parts of the form factors and the self-energy functions in
Eqs.~(\ref{eq:SnI},\ref{eq:GnI},\ref{eq:JnPole}), it would  be inconsistent with  the
K-matrix approach. In particlular, the on-shell contributions $i \delta$
would be taken into account twice for every $\pi N$ propagator
$\mathcal{G}$. An exception are some one-particle irreducible diagrams
contributing to the $\mathcal{T}$-matrix.

It is known that the nucleon and pion degrees of freedom are not enough for
a realistic description of $\pi N$ scattering (for example, the role of both the Delta
resonance and the $\rho$ meson is indispensable)
\cite{gross,pearce,julich,sato,gouds,korsch}.
Since in our model we confine ourselves to the pion and the nucleon, no
calculations for the $\pi N$ scattering observables will be presented in
this paper.

\section{Numerical Results \label{sec:III}}

Two sets of calculations were done, corresponding to the two following
cut-off function $G^0(p^2)$, \eqref{Flam}:
\beq
G^0_I(p^2)=f\left[ \frac{(\lambda^2-m^2)^2}{(\lambda^2-m^2)^2+
(p^2-m^2)^2} \right]^2
\eqlab{23}
\eeq
and
\beq
G^0_{II}(p^2)=f \, e^{-\,\frac{(p^2-m^2)^2}{2 d m^4}}\,.
\eqlab{24}
\eeq
The functional dependence of $G_I^0(p^2)$ is taken from Ref.\ \cite{gross},
where it was used as an off-shell form factor in the $\pi N N $ vertex.
We define a parameter $\Lambda^2=p^2_{0.5}-m^2$, where
$p^2_{0.5}$ is the point at which $G^0(p^2)$
reduces by factor two comparing to its maximum value $f$
(here $p^2_{0.5}>m^2$). Then, for the calculations with the functions
\eqref{23} and \eqref{24}, $\Lambda^2$ equals
\beq
\Lambda^2_{I}(\lambda)=\sqrt{\frac{(\lambda^2-m^2)^2}
{\sqrt{0.5}}-(\lambda^2-m^2)^2}
\eqlab{25}
\eeq
and
\beq
\Lambda^2_{II}(d)=\sqrt{-2 d m^4\,\ln{0.5} }\,,
\eqlab{26}
\eeq
respectively. We find that the iteration procedure described above
converges {\it only} if $\lambda \leq \lambda_c \approx 1.7 \text{GeV}$
for $G_I^0(p^2)$,
and if $d \leq d_c \approx 1.65$ for $G_{II}^0(p^2)$.
The corresponding ``critical" values for the half-widths can be inferred from
\eqref{25} and \eqref{26}: $\Lambda^2_{I}({\lambda}_c)=1.28 \text{GeV}^2$
and $\Lambda^2_{II}(d_c)=1.33 \text{GeV}^2$.
Results of calculations are presented below for the following two cases:
\begin{description}
\item[Case (I)]
Calculations with the cut-off function \eqref{23},
where $\lambda=\lambda_c=1.7 \text{GeV}$;
\item[Case (II)]
Calculations with the cut-off function \eqref{24}, where $d=d_c=1.65$.
\end{description}
As stated above, the constants $f,Z_2$ and
$\delta m$ are chosen to satisfy Eqs.~(\ref{eq:8}). The values
of these constants for cases (I) and (II) are given in Table 1.

The convergence
was considered achieved at iteration m if all the results of
iterations m+1,...,m+20 were
identical to those of iteration m up to six significant digits.
We could put a very strong convergence
criterion since the computer program uses little CPU-time.
With this criterion, convergence was reached after about 100 iterations.
We mention that, for example, the self-energy after 10 iterations differs still
quite noticably from the converged result.

A comparison of results obtained with the two
different cut-off functions show how these reflect in the final
results. The non-perturbative aspects are stressed by comparing the results
of the first iteration (basically a one-loop calculation) with those of the
converged calculation.

\subsection{Results for the half-off-shell form factors}

The imaginary parts of the form factors $G_{PV}(p^2)$ and $G_{PS}(p^2)$ are
shown in Fig.~3. The results of calculations for the two cut-off
functions introduced in \eqref{23} and \eqref{24} are shown next to each
other. For case (II) the tails of the form factors at large off-shellness are
suppressed due to the exponential in the cut-off function. Independent of
the choice of the cut-off function there is a
marked difference in the results of the first iteration (dotted curve) and the
converged results for the pseudovector form factor.
The reason for this
difference is the (small) pseudoscalar component of the final form factor.
The converged and first
iteration results for the pseudoscalar form factor
differ much less, as can be seen from the bottom panels of
Fig.~3.

The real parts of the form factors are shown in Fig.~4. The top pannels show
the pseudovector form factor $G_{PV}(p^2)$ (the solid line) together
with the zeroth iteration form factor $G_{PV}^0(p^2)$ (the dotted line)
which equals the cut-off functions \eqref{23} (left)  and
\eqref{24} (right). It is seen that the bulk of $G_{PV}(p^2)$
is contained already in
$G_{PV}^0(p^2)$, and only a small part comes from the loop
corrections. This manifests itself also in the small difference between the
constant $f$ and the physical coupling constant $g$, as can be read from
Table 1. We conclude that, in the present model, the shape of the converged
form factor $G_{PV}$
depends strongly on the phenomenologically introduced cut-off function. The
middle panels of Fig~4 give more insight in the role of the pion
dressing. There, the real part of the pseudoscalar form factor
$G_{PS}(p^2)$ of the first iteration (the dashed line) is shown together with
the converged result (the solid line). Since the zeroth iteration vertex is
chosen purely pseudovector (\eqref{Flam}), $G_{PS}(p^2)$
appears solely due to the dressing. Also shown is the
difference $Re(G_{PV}(p^2)-G_{PV}^0(p^2))$ which is the dressing
contribution to the real part of the pseudovector form factor (the
dash-dotted and dotted lines for the first iteration and the converged result,
respectively). Note that the deviation of the non-perturbative result form
that of the first iteration is considerable for this quantity.
The ratio of the real parts of the $G_{PS}(p^2)$--
and $G_{PV}(p^2)$-- form factors is given in the bottom panel of Fig~4.
It is small below the pion threshold (e.g.,
$G_{PS}(w_{th}^2)/g$ is about 2.1 \% for both cases (I) and (II)), but
becomes larger at higher $p^2$.
Note that $G_{PV}(p^2)$ decreases for case (II) faster than for case (I),
whereas the behaviour of $G_{PS}(p^2)$ for the two cases is comparable. This
explains why the absolute value of $(Re G_{PS}(p^2))/(Re G_{PV}(p^2))$ grows
faster for case (II) than for case (I).

We remark that admixtures of the pseudovector and pseudoscalar
pion-nucleon couplings have been studied in the past in connection with the
$N N$ and $\pi N$ scattering processes, where the vertex has been determined
by adjusting phenomenological parameters to fit data (see discussions in
%possibly add more references here in line with suggestions of the referee
\cite{grossord,gross,pearce,gouds}). In those calculations the admixture is
assumed to be constant. Instead, the present results indicate that the ratio
is strongly dependent on the
momentum of the off-shell nucleon. Evidence for large pseudoscalar
admixtures for far off-shell momenta has also been observed in
calculations of pion-photoproduction\cite{Kam98}.

\subsection{Results for the self-energy}

The imaginary and real parts of the functions $A(P^2)$ and $B(P^2)$ are shown in
Figs. 5 and 6, respectively.
The solid (dotted) lines are the converged (first iteration) results. One can
see that these functions approach zero
faster for case (II) than for case (I). Of course, this is
entailed by the softer behaviour of $G_{PV}^0(p^2)$ for case (II) as opposed to
case (I) (see Fig.~4).
The difference between
the converged results and those of the first iteration is
substantial, especially  for the function $B(p^2)$.
The on-shell value $\Sigma_L(m)=m(A(m^2)+B(m^2))=Z_2 \delta m$
is negative and equals $-48.8$ MeV for case (I) and $-47.2$ MeV for case
(II), see Table 1. Please note that if we had chosen smaller values for the
cut-off, these self-energy corrections would have been less.
For comparison, we mention that the contribution
to the nucleon mass shift from one-pion loop calculated in
baryon chiral perturbation theory yields a nucleon mass shift
of about $-15$ MeV\cite{meissn}.

Having obtained $A(p^2)$ and $B(p^2)$, one can find the spectral function
of the self-energy $T(\omega)$ from \eqref{30}. Fig.~7 shows
the spectral
function for the two cases (the upper and lower panels). The dotted and
dash-dotted lines are, respectively, the first iteration and the converged
spectral function found in our model. In
spite of the fact that $Im B^1(p^2)$ differs considerably from $Im B(p^2)$,
having the opposite sign at some momenta-squared (see Fig.~5), the spectral
function remains positive for all iterations (as it should).

The spectral function of the nucleon propagator was recently considered in
Ref.\cite{wilets} whose approach is, however,
different from the present work. In particular, there the vertex was
not calculated consistently with the nucleon propagator.

\section{Conclusions}

We have presented a solution procedure for the Schwinger-Dyson equation to
obtain consistently the nucleon self-energy and the half-off-shell
pion-nucleon vertex. Retaining the non-perturbative aspects of the equations
is important. We observe a large difference between the simple one-loop
results and those of the converged procedure. As part of the regularization
procedure, we have introduced a cut off function. We found that in our model
there exists a critical half-width, of the order of $1.3$ GeV$^2$, below which
a non-perturbative solution can be obtained.
Particularly noteworthy is that
even though the dressed vertex near threshold is largely pseudovector in
nature, we find sizable admixtures of
pseudoscalar coupling at large off-shell nucleon momenta.

It is important to realize that off-shell from factors and
self-energies can not be directly measured. The observables
in quantum field theory are obtained from the S-matrix, not
the Green's functions. The latter will depend on the representation of the fields
in the Lagrangian of the theory, whereas the former do not. In principle,
observables should not depend on the regularization
and renormalization procedures chosen. This, however, does not apply to the
Green's functions. To draw qualitative conclusions about physical processes,
one should treat all off-shell ingredients of a model consistently, calculating
them with the same Lagrangian and adhering to the same model assumptions.
For example, the present model for the pion-nucleon off-shell
form factors and the nucleon self-energy can be consistently utilized in a
K-matrix approach to $\pi N$ scattering, as shown in Section III C.

Although the present work deals with $\pi NN$ vertices where
one of the nucleons is off-shell,
we mention here that there exists a large amount of work on the form
factor $G_1(m^2,m^2,q^2)$ (see \eqref{GenV}) for on-shell nucleons and an
off-shell pion (see, e.g., \cite{flender,julon,onsh} and references therein).
In particular, in Ref. \cite{flender} a system of $N$, $\pi$, $\Delta$,
$\rho$, $\epsilon$, and $\omega$ hadrons is considered in a consistent field
theoretical framework. The approach of Ref. \cite{julon}
is based on a meson-exchange model for $\pi N$ scattering, where,
apart from the nucleon and the pion, also Delta isobar and correlated
$\pi \pi$ exchange contributions are included.
As mentioned before, the Delta
resonance and the $\rho$ meson are important ingredients in a
quantitative description of $\pi N$ scattering and pion photoproduction.
Therefore, one should expect these degrees of freedom to play a prominent role
in a realistic model for pion-nucleon form factors.
In the present work, only nucleon and pion fields are included, and
only discontinuities associated with the one-pion threshold
are taken into acoount. In this simplified model we focus on non-perturbative
aspects of the consistent dressing of the pion-nucleon vertex and the
nucleon self-energy. Contributions to the imaginary parts from higher
thresholds can be included in our model either
explicitely, by allowing intermediate states with two or more pions,
or effectively, by considering baryon and meson
resonances (for example, $\Delta$ and $\rho$). This extention of the model
is in progress.

\acknowledgements

This work is part of the research program of the ``Stichting voor
Fundamenteel Onderzoek der Materie'' (FOM) with financial support
from the ``Nederlandse Organisatie voor Wetenschappelijk
Onderzoek'' (NWO). We  would like to thank
Alex Korchin and Rob Timmermans for discussions.

\appendix
\section{The self-energy}

Here some details on the evaluation of the imaginary
part of the nucleon self-energy are given.
To calculate the imaginary parts of the self-energy functions, we need to
evaluate the pole contribution $I_{pole}(p)$ as given in \eqref{InPole}.
In general the integral can be expressed as
$I_{pole}(p)=\gamma_\mu \widetilde{I}^\mu_1(p) + \widetilde{I}_2(p)$,
where $\widetilde{I}^\mu_1$ and $\widetilde{I}_2$
are scalars in spinor space. Since the only
Lorentz vector in the problem is $p^\mu$, $\widetilde{I}^\mu_1$
must be proportional to
it, $\widetilde{I}^\mu_1(p) = ((\widetilde{I}_1 \cdot p )/p^2) \,p^\mu$.
Hence, one may write $I_{pole}(p) = \vslash{p} I_1(p^2) + I_2(p^2)$, where
$I_1(p^2)=(\widetilde{I}_1 \cdot p )/p^2$ and $I_2(p^2)=\widetilde{I}_2(p^2)$
are Lorentz scalars. They equal
\bea
I_1(p^2) &=&
\frac{p^2+m^2-\mu^2}{32 \pi p^4} \, r(p^2) \, \theta(p^2-(m+\mu)^2)\\
I_2(p^2) &=& \frac{m}{16 \pi p^2} \,r(p^2)\, \theta(p^2-(m+\mu)^2).
\eea
where $r(p^2)=\sqrt{\lambda(p^2,m^2,\mu^2)}$, with the K\"all\'en function
defined as $\lambda(x,y,z)\equiv (x-y-z)^2-4yz$.
Using these expressions and introducing
the shorthand notation, $g_{1,2} \equiv Re G_{1,2}^n(p^2)$,
the imaginary parts of the self-energy functions can
be determined from Eqs.~(\ref{eq:Sdrs},\ref{eq:SnI},\ref{eq:3}),
\begin{eqnarray}
Im A^{n+1}(p^2)&=&
3\Big(- (g_1-g_2)^2 I_1(p^2) + 2 (g_1-g_2)g_2 {I_2(p^2) \over m} -
g_2^2 \frac{p^2 I_1(p^2)}{m^2} \Big)  \nonumber \\
&=&-\frac{3}{32 \pi p^2}\,r(p^2)\,
\theta(p^2-(m+\mu)^2)\,\Big\{
\frac{p^2+m^2-\mu^2}{p^2} g_1^2 \nonumber \\  &&
+\big(\frac{5 p^2+m^2-\mu^2}{p^2}+
\frac{p^2+m^2-\mu^2}{m^2}\big) g_2^2-
2 \frac{3p^2+m^2-\mu^2}{p^2} g_1 g_2\Big\}\,,\\
Im B^{n+1}(p^2)&=&
3\Big((g_1-g_2)^2 {I_2(p^2) \over m} - 2 (g_1-g_2)g_2
{p^2 I_1(p^2) \over m^2} +
g_2^2 {p^2 I_2(p^2) \over m^3} \Big) \nonumber \\
&=&\frac{3}{16 \pi p^2}\,r(p^2)\,
\theta(p^2-(m+\mu)^2) \nonumber \\  && \times\Big\{
g_1^2+\frac{2 p^2+2 m^2-\mu^2}{m^2} g_2^2-
\frac{p^2+3m^2-\mu^2}{m^2} g_1 g_2\Big\}\,,
\end{eqnarray}
the factor 3 in the above equations results from the multiplication
of the isospin matrices, $\tau_{\alpha}\tau_{\alpha}=3$,
and the minus sign in front of
$I_1(p^2)$  from commuting the $\gamma^5$ matrices.
The real parts of
$A^{n+1}(p^2)$ and $B^{n+1}(p^2)$ are found by applying dispersion relations
\eqref{dispA}, where all integrals are done numerically.

For later use, the dressed nucleon propagator is written as
\beq
\big( S(p) \big)^{-1} = Z_2(\vslash{p}-m_B)
-\Big(Re A(p^2) \vslash{p} + Re B(p^2) m \Big)
=\alpha(p^2) \Big( \vslash{p}-\xi(p^2) \Big),
\eqlab{A6}
\eeq
where
\beq
\alpha(p^2)=Z_2-Re A(p^2)\,,\;\;\;\; \xi(p^2)=\frac{Z_2\,(m-\delta m)+
Re B(p^2)\, m}{\alpha(p^2)}.
\eqlab{A7}
\eeq
The spectral function of the self-energy is introduced through
\beq
\Sigma_L(p)=\zeta(\omega)\Lambda^{+}(\vslash{p})+
\zeta(-\omega)\Lambda^{-}(\vslash{p})\,,
\eqlab{29}
\eeq
where $\Lambda^{\pm}(\vslash{p})=(\pm \vslash{p}+\omega)/(2\omega)$
are the projectors on
positive- and negative-energy states of the nucleon with
the invariant mass $\omega=\sqrt{p^2}>0$. The spectral function can now be
defined as\cite{bogol}
\beq
T(\pm \omega)=\mp \frac{1}{\pi}\,Im\,\zeta(\pm \omega)\,.
\eqlab{33}
\eeq
Equating the right-hand side of \eqref{29} and the form of $\Sigma_L(p)$
from \eqref{Sdrs} yields
\beq
T(\pm \omega)=-\frac{1}{\pi} \big(\omega \,Im\, A(p^2) \pm m \,Im\,B(p^2)
\big)\,.
\eqlab{30}
\eeq

\section{The form factors}

The calculation of the imaginary parts of the form factors can be reduced
to computing one-dimensional integrals which are done numerically.
First consider the integral on the right-hand side of \eqref{JnPole}.
$J_{pole}$ can be split as
$J_{pole}=\gamma_{\mu} \widetilde{J}_1^{\mu} + \widetilde{J}_2$, where
$\widetilde{J}_1^{\mu}$ and $\widetilde{J}_2$ are scalars in spinor space,
and a possible rank-2 tensor structure vanishes since $\overline{u}(p')
\vslash{p}' = \overline{u}(p') m$. For the same reason
$\widetilde{J}_1^{\mu}$ is proportional to only the vector $p^\mu$.
Following the same argumentation as used in Appendix A, we can write
\beq
J_{pole}= \vslash{p} J_1 + J_2\,,
\eqlab{Jsplit}
\eeq
where we have introduced the Lorentz scalars $J_1$ and $J_2$.
To write down expressions for $J_1$ and $J_2$, we define the following
functionals:
\bea
K_1[f] &\equiv& \int_{-1}^{1}\!dx\,\frac{f(w^{\prime 2})}{\alpha(w^{\prime 2})},
\eqlab{b4}\\
K_2[f]&\equiv& \int_{-1}^{1}\!dx\,x\,\frac{f(w^{\prime 2})}{\alpha(w^{\prime 2})},
\eqlab{b5}\\
K_3[f]&\equiv& \int_{-1}^{1}\!dx\,\frac{f(w^{\prime 2})}{\alpha(w^{\prime 2})
(w^{\prime 2}-\xi^2(w^{\prime 2}))},
\eqlab{b6}\\
K_4[f]&\equiv& \int_{-1}^{1}\!dx\,x\,\frac{f(w^{\prime 2})}{\alpha(w^{\prime 2})
(w^{\prime 2}-\xi^2(w^{\prime 2}))},
\eqlab{b7}
\eea
where $f$ is any function for which the integrals exist,
$\alpha$ and $\xi$ are given by
\eqref{A7} for the (n+1)$^{\rm st}$ iteration, and
\beq
w^{\prime 2}=(p^{\prime}+k)^2=m^2+\mu^2-\frac{p^4-(m^2-\mu^2)^2}{2p^2}-
\frac{r(p^2)^2}{2p^2}\,x\,,
\eqlab{w}
\eeq
with $x$ being the cosine of the polar angle between the three vectors
$\overrightarrow{p}^{\prime}$ and $\overrightarrow{k}$.
Now $J_1$ and $J_2$ can be written as
\begin{eqnarray}
&&J_1\equiv J_1(p^2)=-\left\{ \Big( K_1-K_2 \Big)
\Big[ \frac{p^2+m^2-\mu^2}{2p^2}\big( g_1g_2+\frac{\xi-4m}{2m}g_2^2 \big)
+\frac{g_2^2}{2} \Big] \right. \nonumber \\
&&\left. + \Big( K_3-K_4 \Big)
\Big[ \big( \frac{p^2+m^2-\mu^2}{p^2}\,\frac{\xi}{4m}-\frac{m^2-\mu^2}{2p^2}
\big) \big( mg_1+(\xi-m)g_2 \big)^2 \Big] \right\} \nonumber \\
&& \times \frac{r(p^2)}{8 \pi p^2}\,\theta(p^2-(m+\mu)^2)\,
\eqlab{b8}
\end{eqnarray}
and
\begin{eqnarray}
&&J_2\equiv J_2(p^2)=-\left\{ \Big( K_1+K_2 \Big)
\Big[ \big( \frac{\xi-4m}{2m}+\frac{p^2+m^2-\mu^2}{4m^2} \big)g_2^2+
g_1g_2 \Big] \right. \nonumber \\
&&\left. +\Big( K_3+K_4 \Big)
\Big[ \big( \frac{\xi-2m}{2m}+\frac{p^2+m^2-\mu^2}{4m^2} \big)
\big( mg_1+(\xi-m)g_2 \big)^2 \Big] \right\} \nonumber \\
&& \times \frac{r(p^2)}{8 \pi p^2}\,\theta(p^2-(m+\mu)^2)\,.
\eqlab{b9}
\end{eqnarray}
Since for a given $f$ in Eqs.~(\ref{eq:b4} - \ref{eq:b7})
the $K_i$ are functions of $p^2$ only, Eqs.~(\ref{eq:b8},\ref{eq:b9},\ref{eq:Jsplit})
show that $J_{pole}$ depends only on the
Lorentz vector $p$ and does not depend on $p'$ as it might appear from
the right-hand side of \eqref{JnPole}.

Using Eqs.~(\ref{eq:Jsplit},\ref{eq:GnI},\ref{eq:3}),
one obtains for the imaginary parts of the form factors:
\begin{eqnarray}
Im G_1^{n+1}(p^2)&=&g_1 J_2(p^2)+
\big(\frac{m^2-p^2}{m^2}g_2-g_1\big)J_1(p^2)
\end{eqnarray}
and
\begin{eqnarray}
Im G_2^{n+1}(p^2)&=&g_2 J_2(p^2)+
\big(g_2-g_1\big)J_1(p^2)\,,
\end{eqnarray}
where $J_1$ and $J_2$ are given by Eqs.~(\ref{eq:b8}) and (\ref{eq:b9}).
Finally, \eqref{dispG} is applied to obtain the real parts of the (n+1)$^{\rm
st}$ iteration for the form factors.

\begin{table}
\caption[1]{ Values of the renormalization constants $f,\,Z_2$ and $\delta m$ for
the two choices for the function $G^0(p^2)$, \eqref{23} and \eqref{24}.}
\begin{tabular}{c|ccc}
\ \ Case\ \  & $f$ & $Z_2$ &$\delta m$ [MeV]\\
\tableline
 ({I}) &12.42 & 0.848 &$-57.4$ \\
 ({II}) &12.43 & 0.848 &$-55.5$ \\
\end{tabular}
\label{tab:1}
\end{table}

\newpage
\begin{figure}
\epsfxsize 12 cm
\centerline{\epsffile[0 234 594 559]{disk$1:[kondrat.pictures]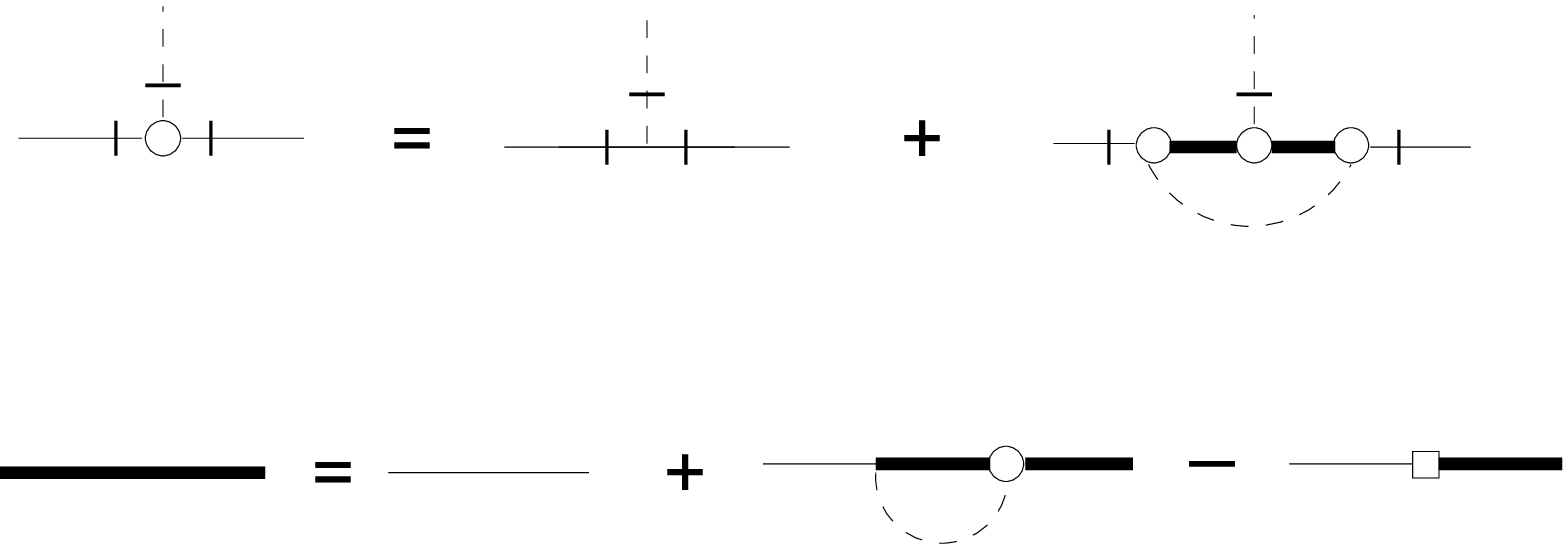}}
\caption[f1]{
The graphical representation of the system of Eqs.~(\ref{eq:sys}).
The thin and thick solid lines correspond to
the free and dressed propagators of the nucleon, respectively. The dashed
line is the propagator of the pion. The circle indicates
the dressed $\pi N N$ vertex, and the square stands for the counterterm
contribution to the nucleon self-energy. In the equation for the vertex the
propagators of the external lines are stripped away, as indicated by the
dashes on these lines.
\label{fig1}}
\end{figure}

\begin{figure}
\epsfxsize 12 cm
\centerline{\epsffile[0 245 593 540]{disk$1:[kondrat.pictures]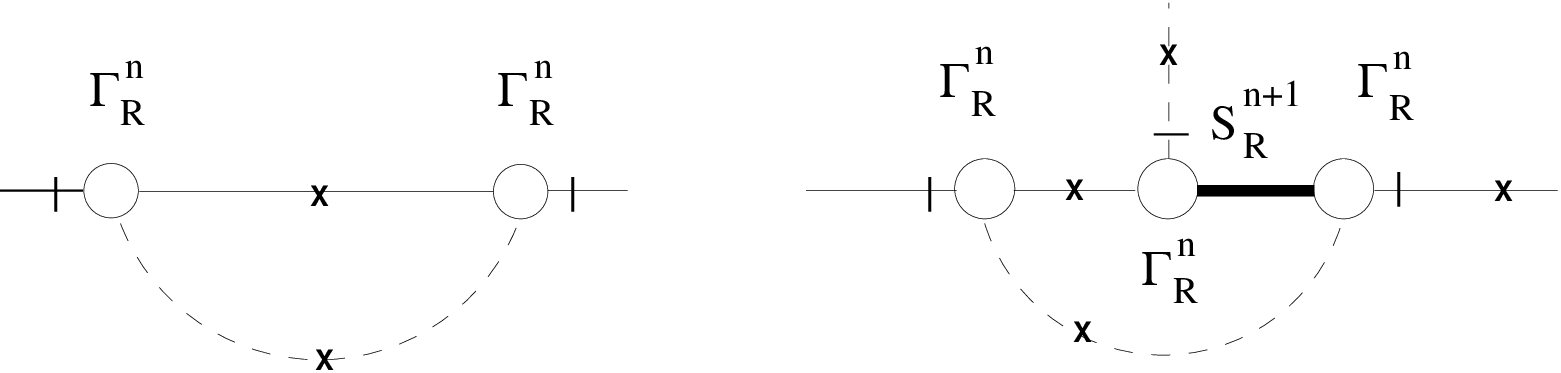}}
\caption[f2]{
The pole contribution of iteration $n+1$ to
the self-energy (the left picture) and
the vertex (the right picture), as expressed by
Eqs.~(\ref{eq:SnI}-\ref{eq:JnPole}). The notation
is as in Fig. 1, with the subscript $R$ indicating that the vertex and
propagator are calculated using only the real parts of the form factors and
self-energy functions. The crosses
on lines indicate that the corresponding particles are put on their mass shells.
\figlab{f2}}
\end{figure}

\begin{figure}
\epsfysize 20 cm
\centerline{\epsffile[0 0 594 841]{disk$1:[kondrat.pictures]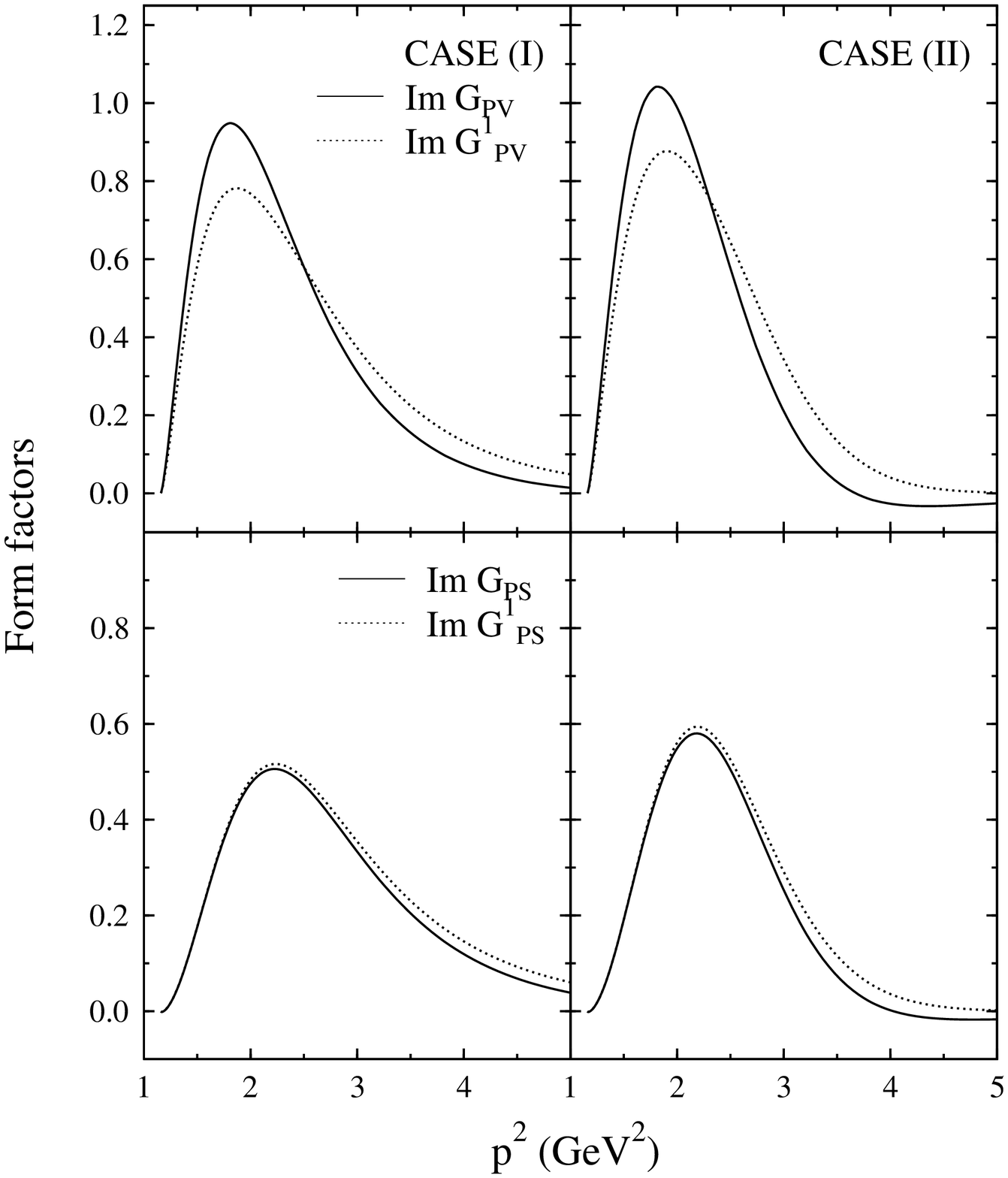}}
\caption[f3]{The imaginary parts of the pseudovector and pseudoscalar
$\pi N N$ form factors as functions of the momentum-squared of the off-shell
nucleon, defined in \eqref{4}. The calculations correspond to the two
cut-off functions \eqref{23} (the left panels) and \eqref{24}
(the right panels). The drawn (resp. dotted) curves show the converged
(resp. first iteration) results.
\label{fig3}}
\end{figure}

\begin{figure}
\epsfysize 20 cm
\centerline{\epsffile[0 0 594 841]{disk$1:[kondrat.pictures]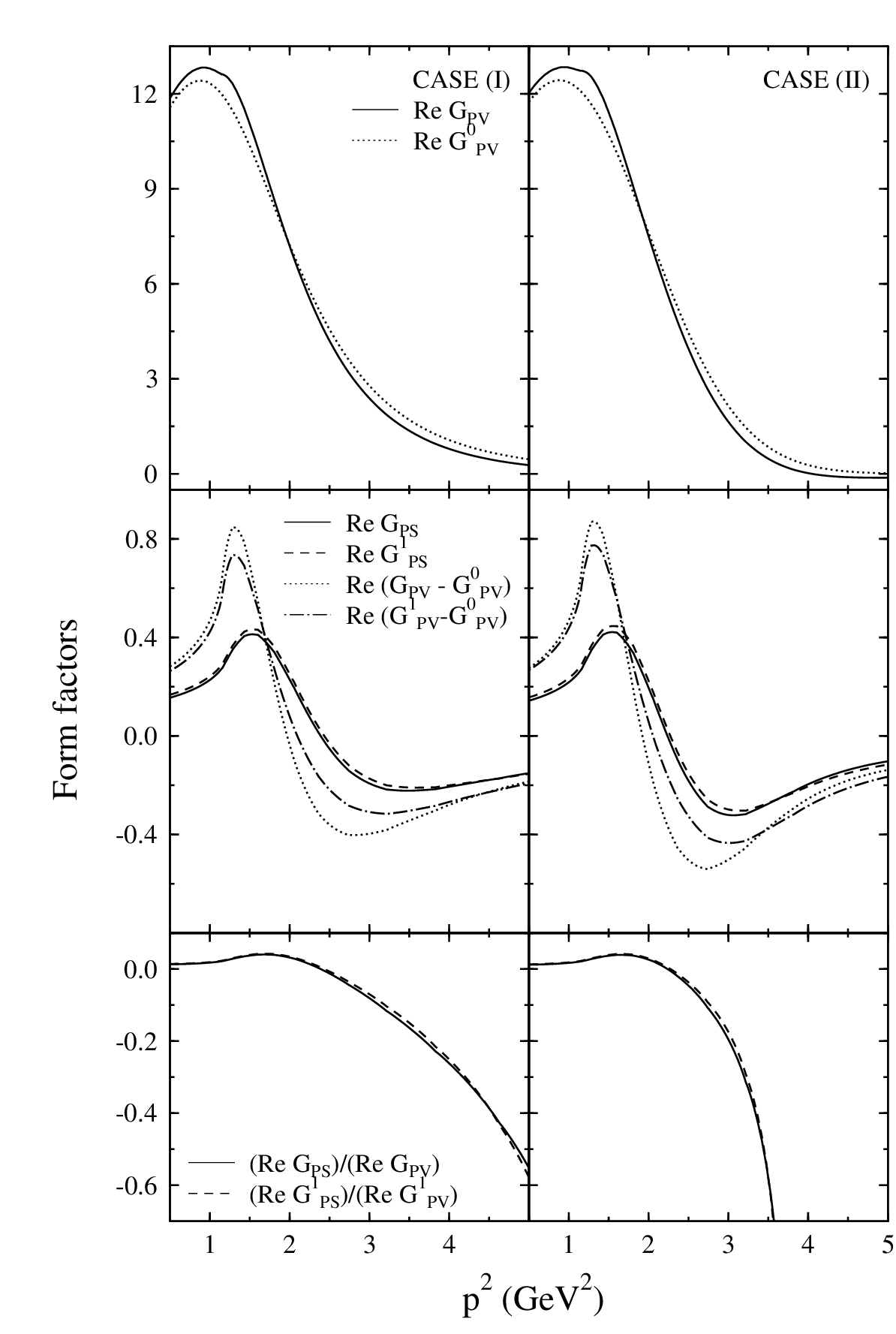}}
\caption[f4]{The real parts of the pseudovector and pseudoscalar
$\pi N N$ form factors as functions of the momentum-squared of the off-shell
nucleon, defined in \eqref{4}. The calculations correspond to the two
cut-off functions \eqref{23} (the left panels) and \eqref{24}
(the right panels).
In the top panels the zeroth iteration and the converged
form factors are given by the dotted and drawn lines, respectively.
In the middle panels the converged results and those of the first iteration
are shown for the pseudoscalar form factor
and the loop contribution to the pseudovector form factor.
The bottom panels show the ratio of the pseudoscalar and pseudovector form
factors, where the drawn (resp. dashed) curves correspond
to the converged (resp. first iteration) results.
\label{fig4}}
\end{figure}

\begin{figure}
\epsfysize 10 cm
\centerline{\epsffile[0 55 594 585]{disk$1:[kondrat.pictures]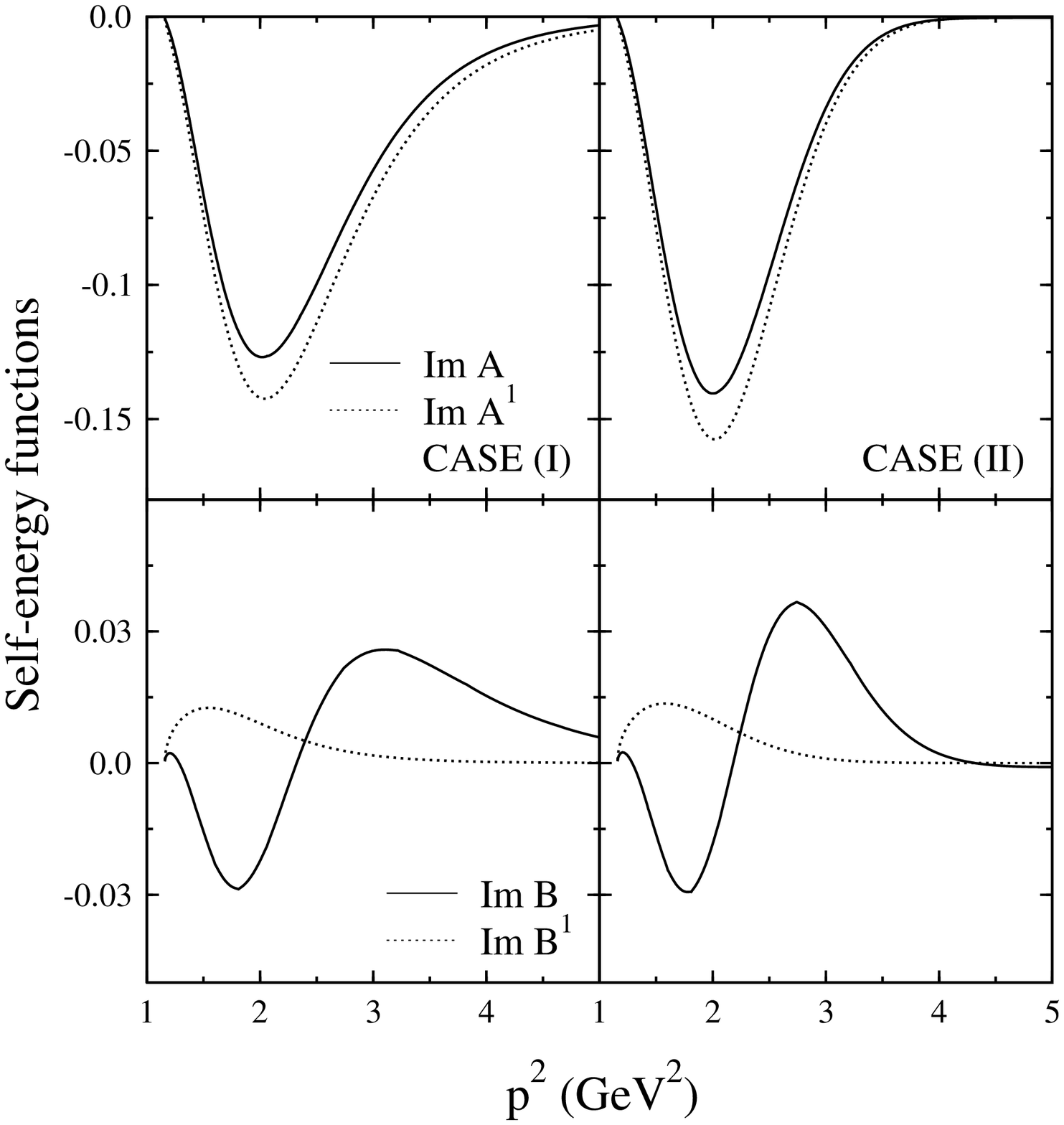}}
\caption[f5]{The imaginary parts of the self-energy functions
$A(p^2)$ and $B(p^2)$, as defined in \eqref{Sdrs}. The calculations
correspond to the two cut-off functions \eqref{23} (the left panels)
and \eqref{24} (the right panels).
The drawn (resp. dotted) curves are the converged (resp. first iteration)
results.
\label{fig5}}
\end{figure}

\begin{figure}
\epsfysize 10 cm
\centerline{\epsffile[0 55 594 600]{disk$1:[kondrat.pictures]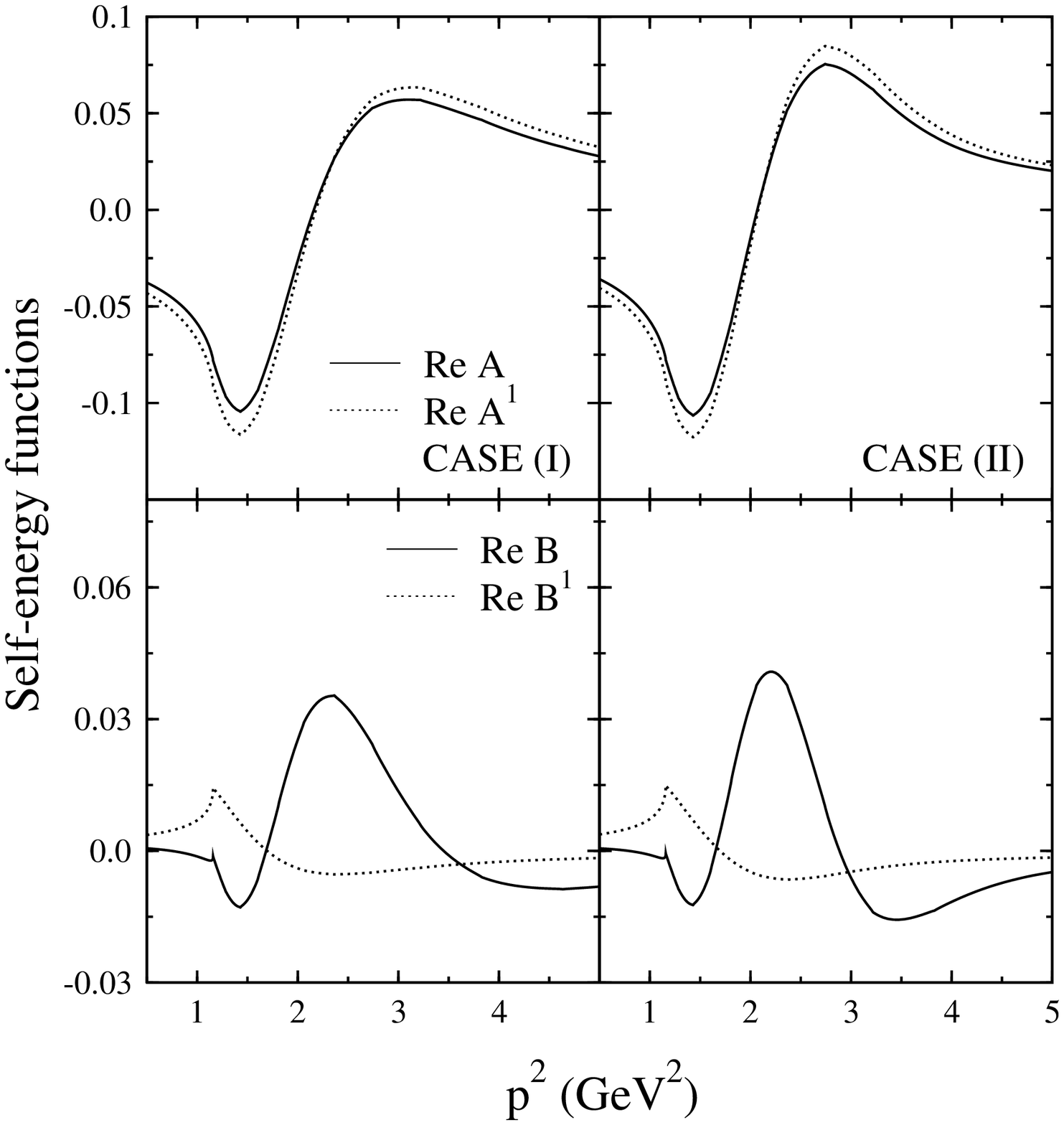}}
\caption[f6]{The same as in Fig. 5, but for the real parts of
$A(p^2)$ and $B(p^2)$.
\label{fig6}}
\end{figure}

\begin{figure}
\epsfysize 20 cm
\centerline{\epsffile[0 50 594 770]{disk$1:[kondrat.pictures]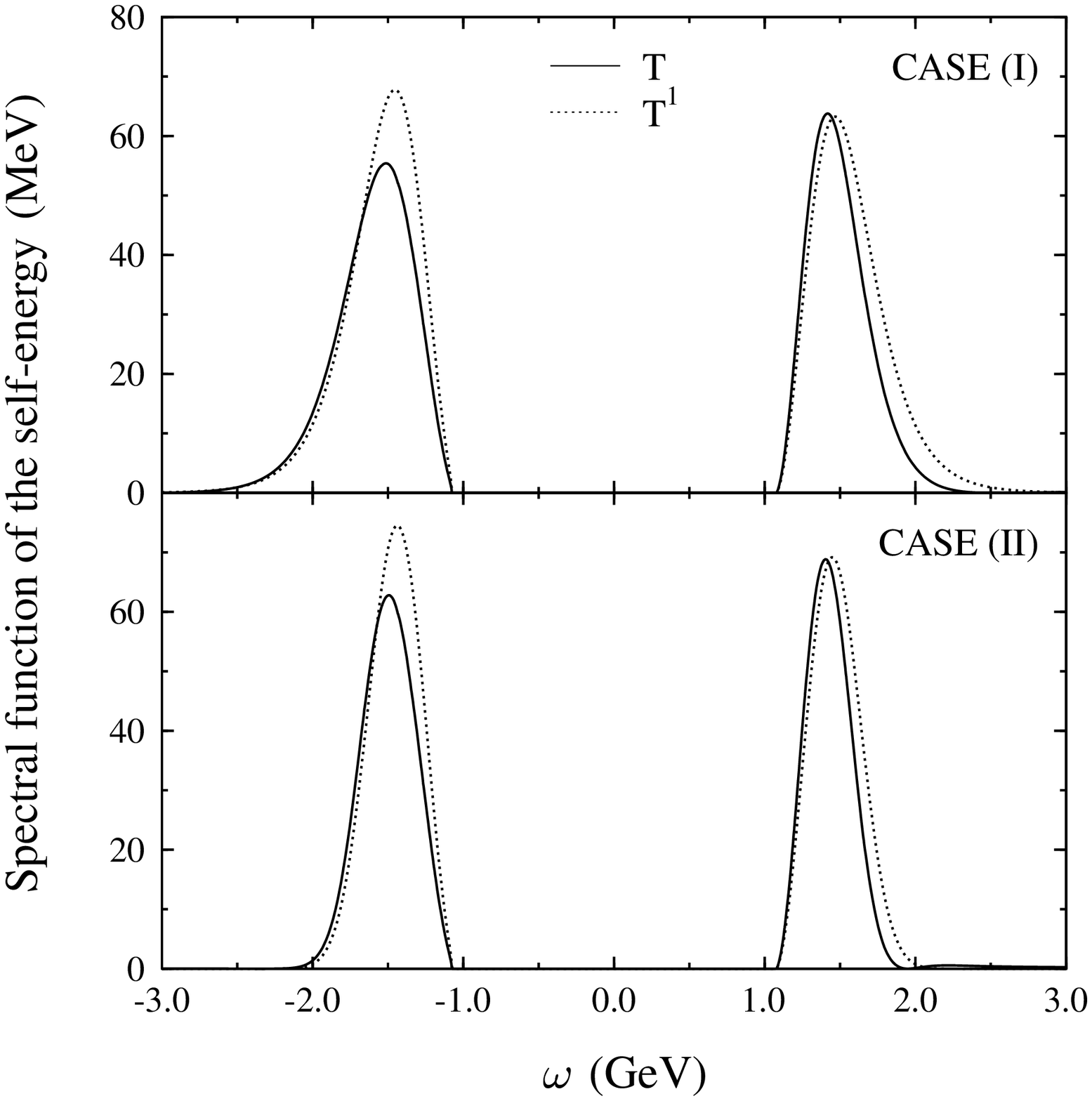}}
\caption[f7]{The self-energy spectral function
$T(\omega)$ as function of invariant mass of the nucleon.
The upper (resp. lower) panel
corresponds to the calculations with the cut-off function \eqref{23}
(resp. \eqref{24}). The drawn (resp. dashed) curves are the converged
(resp. first iteration) results.
\label{fig7}}
\end{figure}

\end{document}